\documentstyle[twocolumn,aps]{revtex}
\begin{document}

\twocolumn[\hsize\textwidth\columnwidth\hsize\csname
@twocolumnfalse\endcsname

\title{
Ultrafast Dynamics of Interfacial Electric Fields in Semiconductor 
Heterostructures  Monitored by Pump-Probe Second Harmonic Generation
}
\author{Yu. D. Glinka, T. V. Shahbazyan, I. E. Perakis and N. H. Tolk}
\address{Department of Physics and Astronomy, Vanderbilt University,
  Nashville, TN 37235}   
\author{X. Liu, Y. Sasaki and J. K. Furdyna}
\address{Department of Physics, University of Notre Dame, Notre Dame,
IN 46556}
%\date{\today}
\maketitle
\draft
%\wideabs
\begin{abstract}
We report first measurements of the ultrafast dynamics of interfacial
electric fields in semiconductor multilayers using pump-probe second
harmonic generation (SHG). A pump beam was tuned to excite carriers in
all layers of GaAs/GaSb and GaAs/GaSb/InAs heterostructures. Further
carrier dynamics manifests itself via electric fields created by by
charge separation at interfaces. The evolution of interfacial fields
is monitored by a probe beam through the eletric-field-induced SHG
signal.  We distinguish between several stages of dynamics originating
from redistribution of carriers between the layers. We also find a
strong enhancement of the induced electric field caused by
hybridization of the conduction and valence bands at the GaSb/InAs
interface.
\end{abstract}
%\pacs{PACS numbers: 78.47.+p, 72.20.Jv, 73.40.Kp}

\vskip2pc]

The ultrafast dynamics of optically excited carriers in semiconductors
is an important issue in solid-state physics\cite{chemla01,perakis00}.
The knowledge of processes governing the carrier relaxation in solids
is essential for designing novel multifunctional high-speed electronic
and optoelectronic devices\cite{othonos98}. This is particularly
important for MBE grown multilayer semiconductor structures which
offer the greatest promise for nanoscale electronics. Quantum
confinement is known to significantly affect the carrier
thermalization dynamics in quantum wells\cite{bolton98} and quantum
dots.\cite{borri01} On the other hand, one should expect that the
interface between different semiconductors will influence the carrier
dynamics even in non-quantum-confined structures. This is especially
the case when the band offsets of adjacent semiconductor layers are
significant.  During the fast relaxation due to electron-electron 
($< 200$ fs) and electron-phonon ($\sim 1$ ps) scattering
processes\cite{othonos98,bolton98,borri01,lin87}, the electrons and
holes tend to accumulate at different sides of the interface.  The
resulting charge separation gives rise to the interfacial electric
fields which can, in principle, change the initial band alignment and
give rise to interlayer transport phenomena\cite{koshihara97}, In such
situations, the adequate description of carrier dynamics should take
into account both optically-induced interfacial electric fields and
carrier population.

The techniques typically used for monitoring the ultrafast carrier
dynamics in semiconductors are pump-probe transmission or reflection
spectroscopy\cite{chemla01,perakis00,othonos98,bolton98,borri01,lin87},
These methods rely on the linear response of the electron-hole
subsystem, excited by pump pulse, to the probing light. As a result,
they give an accurate account of the carrier population dynamics,
while being insensitive to the interfacial fields. On the other hand,
{\em nonlinear} (in probing light) optical methods such as second
harmonic generation (SHG) are known to be extremely sensitive to
electric fields occurring at the surfaces and
interfaces\cite{shen84,lupke99}. This unique feature of SHG was
employed to study the carrier relaxation dynamics at the silicon-oxide
interface\cite{lupke99,downer99,glinka02}. In particular, the excited
electrons and holes were trapped at different sides of the interface
and the resulting long-time dynamics (of the order of minutes) was
monitored by the electric-field-induced SHG (EFISHG) signal. Note that
in such systems, the electric fields are fairly uniform and are well
described within dipole approximation\cite{downer99}.

In this paper we report the first measurements of the {\em ultrafast} dynamics
of the optically-induced interfacial electric fields in semiconductor
multilayers using time-resolved pump-probe SHG spectroscopy. The measurements
were performed on GaAs/GaSb and GaAs/GaSb/InAs heterostructures.  The crucial
advantage of our technique as compared to the usual pump-probe spectroscopy
lies in its sensitivity to {\em spatially separated} regions of
heterostructures. Therefore, the pump-probe SHG allows us to 
{\em simultaneously} monitor the dynamics at different interfaces of the same
sample. In particular, by tracking the evolution of different
interfacial fields contributing to total EFISHG signal we are able
to study the redistribution of carriers between the layers due to
electron transport across interfaces. Note that the
carrier transverse diffusion on GaAs surface plane was demonstrated previously
in pump-probe spatially separated SHG experiment\cite{qi95}. However, to the
best of our knowledge, the optical detection of {\em interlayer} electron
transport in semiconductor heterostructures has not been reported before.

The GaSb/InAs spatial interface is of special interest because of an unusual
hybridization of the conduction InAs and valence GaSb band states\cite{meyer}.
In semimetallic samples, the effect of band mixing was observed in an
appearance of energy gaps in capacitance-voltage and quantum Hall
measurements\cite{yang97} as well as in a splitting of cyclotron resonance
peak measured using far-infrared
spectroscopy\cite{kono97,marlow99,comanescu02}. In our experiment, we find
that the band hybridization results in a strong enhancement of the EFISHG
signal.
                
We have investigated four heterostructures grown by molecular beam epitaxy:
(1) GaAs/GaSb(20 nm); (2) GaAs/GaSb(400 nm); 
(3) GaAs/GaSb(500 nm)/InAs(20 nm), with an InSb interface between GaSb and
InAs layers; and (4) GaAs/GaSb(500 nm)/InAs(20 nm), with a GaAs interface
between GaSb and InAs. In 
all samples, the thickness of GaAs layer was 100 nm.  Samples were grown on
semi-insulating (100) GaAs substrates.  Prior to GaSb or GaSb/InAs deposition
the substrates were cleaned in situ by oxide desorption, by heating to 600
$^{\circ}$C, after which a GaAs buffer layer was grown at 590
$^{\circ}$C. In deposition of the overlayers, we used 490 $^{\circ}$C
for GaSb growth and 450$^{\circ}$C for InAs.

A pump-probe configuration with linearly-polarized pump and probe beams was
used in our measurements. The observed EFISHG signal was monitored as a
function of probe-to-pump delay times.  All optical measurements were carried
out in a liquid helium cryostat at 4.3 K.  The initial beam of 150-fs pulses
from a mode-locked Ti:Al2O3 laser (Mira 900) at the wavelength of 800 nm (1.55
eV) and a repetition rate of 76 MHz was split into pump and probe beams.  The
probe beam of 120-mW-average power has passed through an optical delay stage.
The pump beam, after being chopped at a frequency of 400 Hz, was of the same
average power. The overlap spot of the pump and probe beams on the sample was
100 $\mu m$ in diameter.  The pump beam was incident normally on the sample
with either $p$ or $s$ polarization. The probe beam (also $p$ or $s$
polarized) was directed to the sample surface at the angle of 75$^{\circ}$.
The SHG signal was optically separated from the reflected fundamental probe
beam and measured by a photomultiplier tube through a ``lock-in'' amplifier
triggered by the chopped pump pulses.

Figures 1 and 2 show pump-induced SHG signals measured for a GaAs/GaSb
heterostructure (sample 2) in comparison with those taken for GaAs/GaSb/InAs
with an InSb interface (samples 3) and with a GaAs interface (sample 4),
respectively.  Note that the signal is observed only for $p$-polarized probe
beam.  Switching the pump polarization from $p$ to $s$ does not affect
significantly neither the time dependence of signals nor their intensities.
No pump-induced SHG signal was observed for GaAs/GaSb sample with thinner 
(20 nm) GaSb layer (sample 1).

The measured signals were fitted by a combined exponential rise/decay function
as shown in Figs. 1 and 2 by solid lines. According to the fit, the
pump-induced SHG signals taken with either $p$ or $s$ pump polarization for
GaAs/GaSb heterostructure (sample 2) is described by a single rise-time
constant $\tau_{R_1}\simeq 2 (\pm 1)$ ps and two decay-time constants
$\tau_{D_1}\simeq 15 (\pm 3)$ ps and $\tau_{D_2}\simeq 100 (\pm 10)$ ps.  In
sharp contrast, the evolution of the pump-induced SHG signal for
GaAs/GaSb/InAs heterostructures shows {\em two} stages in the signal rise. The
fast-rising component is characterized by the same rise-time constant as that
measured for GaAs/GaSb sample ($\tau_{R_1} \sim 2$ ps). The additional
slower-rising component in both GaAs/GaSb/InAs samples has rise-time constant
$\tau_{R_1} \simeq 10 (\pm2)$ ps which is slightly smaller than $\tau_{D_1}$.
At the same time, the decay-time constants for GaAs/GaSb/InAs and  GaAs/GaSb
samples are similar.

The intensity of pump-induced SHG signals measured for GaAs/GaSb/InAs
heterostructures shows strong dependence on the interface type between GaSb
and InAs layers. For sample 3 (with InSb interface), the signal amplitude is
significantly larger than for GaAs/GaSb structure (Fig. 1) while for sample 4
(with GaAs interface) the signal is comparable to that for GaAs/GaSb (Fig. 2).
Moreover, despite similar decay-time constants for both GaAs/GaSb/InAs
samples, there is considerable long-time (delay-time $> 250$ ps) constant
background for sample 3 while it is significantly smaller for sample 4.

An apparent presence of several stages in the evolution of measured SHG signal
together with its sensitivity to the interface type of GaAs/GaSb/InAs
heterostructures indicate a rather complex dynamics of interfacial electric
fields originating from a redistribution of carriers between the interfaces.
The induced {\em local} electric fields, ${\cal E}_{1}(t)$ and 
${\cal E}_{2}(t)$ (subscripts $1$ and $2$ refer to GaAs/GaSb and GaSb/InAs
interfaces, respectively), depend on the number of carriers as well as on their
spatial distribution near each interface at a given time. The observed EFISHG
signal is determined by contributions from {\em all} interfacial electric
fields.  Retaining only linear terms in ${\cal E}_{i}(t)$, the nonlinear
polarization can be written as \cite{shen84,lupke99,glinka02}
\begin{equation}
\label{pol}
P^{NL}(2\omega,t) =\Bigl[\chi^{(2)}+\chi_{1}^{(3)}{\cal E}_{1}(t)
+\chi_{2}^{(3)}{\cal E}_{2}(t)
%+\chi_I^{(Q)} \nabla_z{\cal E}_I(t)\bigr]
\Bigr][E(\omega)]^2,
\end{equation}
where $E(\omega)$ is the electric field component of the incident probe light,
$\chi^{(2)}$ is the second-order bulk susceptibility, and $\chi_{i}^{(3)}$ are
the third-order susceptibilities at the interfaces.  In general, a third-order
nonlinear polarization $P^{(3)}(2\omega,t)$ at the frequency $2\omega$ is
generated only when the induced field ${\cal E}_{i}(t)$ and the optical field
$E(\omega)$ have the same polarization. In our case, only $p$-polarized light
exhibits an EFISHG signal since the interfacial electric fields are directed
along the normal to the interface.  In fact, the EFISHG signal is not very
sensitive to the pump polarization because the interfacial fields arise as a
result of {\em incoherent} relaxation of carriers. Note that the latter is the
reason for the absence of signal at negative delay-times which originates from
coherent effects in the population dynamics\cite{bartels97}. The measured
EFISHG intensity, $\Delta I^{(2\omega)}=I^{(2\omega)}-I_0^{(2\omega)}$, which
is obtained by subtracting the bulk contribution
$I_0^{(2\omega)}=|\chi^{(2)}|^2|E(\omega)|^4$ from the total intensity,
$I^{(2\omega)}=|P^{NL}(2\omega)|^2$, has the following form (after neglecting
higher-order nonlinear in ${\cal E}_{i}(t)$ terms)
\begin{equation}
\label{int}
\Delta I^{(2\omega)} \propto 
\chi_{1}^{(3)}{\cal E}_{1}(t)
+\chi_{2}^{(3)}{\cal E}_{2}(t)
%\sum_I\chi_I^{(3)}{\cal E}_I(t).
% \chi_{GaAs/GaSb}^{(3)}{\cal E}_{GaAs/GaSb}(t)
% +\chi_{GaSb/InAs}^{(3)}{\cal E}_{GaSb/InAs}(t)
\end{equation}
The third-order susceptibilities $\chi_{i}^{(3)}$ are determined by the energy
spectrum of carriers near the two interfaces, GaAs/GaSb and GaSb/InAs.
Usually, the electrons and holes at the interface are separated by several
atomic layers and, in this case, the susceptibilities for the corresponding
bulk material can be used with a good accuracy\cite{lupke99}. However, for the
GaSb/InAs interface with InSb bonding, the change in the energy dispersion at
the interface vicinity is found to have a significant effect.

We attribute the observed several stages of dynamics of interfacial electric
fields to an interplay between relaxation of carriers and their transport across
heterostructures. Because the laser light was tuned just above the GaAs bandgap,
the electrons were excited in all the layers of heterostructures [inset in
Fig. 2(a)]. Electrons with high excess energies relax to the lower-energy
conduction band states in the GaAs and InAs layers, while the holes are
accumulated in the GaSb layer. The resulting charge separation leads to an
appearance of electric fields across the interfaces. The rise of interfacial
fields manifests itself in the initial growth of the EFISHG signal
characterized by fast rise-time constant $\tau_{R_1}\sim 2$ ps.  Note here
that in sample 1 with thinner (20 nm) GaSb layer, the carriers are accumulated
predominantly in the GaAs layer, so the interfacial fields are weak and the
corresponding EFISHG signal is undetectable.  In all other samples, the
majority of carriers are excited in the thickest GaSb layer leading to
significant concentration of holes in that layer.  The induced interfacial
fields bend the initial energy profile and, in particular, lower the barrier
at the GaAs/GaSb interface [inset in Fig. 2(b)]. As the negative charges at
the GaAs side start to transfer through the barrier, the electric field at the
GaAs/GaSb interface decreases. For GaAs/GaSb heterostructure (sample 2), such
a decrease shows up in a fast decay of the EFISHG signal with decay-time
constant $\tau_{D_1}\sim 15$ ps.  The subsequent relaxation of the interfacial
electric field, characterized by long decay-time constant $\tau_{D_2}\sim 100$
ps, is due to carrier migration away from the interface.

For GaAs/GaSb/InAs heterostructures, the situation is completely different.
In this case, the electrons crossing the GaAs/GaSb interface accumulate in the
InAs layer. The resulting increase in the GaSb/InAs interfacial field
manifests itself as the additional rise component of the EFISHG signal with
rise-time constant $\tau_{R_2}\sim 10$ ps. Note that this arrival time is
comparable to the departure time in the GaAs/GaSb sample.
A subsequent relaxation of the interfacial electric fields is characterized by
similar $\tau_{D_2}\sim 100$ ps decay-time constant.

A striking feature observed for GaAs/GaSb/InAs heterostructures is a
significantly more intense EFISHG signal for InSb type interface between GaSb
and InAs layers than that for GaAs type. We attribute this difference to a
larger overlap between InAs conduction and GaSb valence band envelope
functions across the spatial GaSb/InAs interface with InSb bonds\cite{meyer}.
In this case, the nonlinear susceptibility $\chi^{(3)}$ can no longer be
approximated by a corresponding bulk expression.  The hybridization of
conduction and valence bands leads to anticrossing of bands dispersions which
was observed, in linear absorption, as satellites of the cyclotron resonance
peak.\cite{kono97,marlow99,comanescu02} Importantly, the band splitting also
gives rise to low-energy transitions across the hybridization gap which
enhance the {\em nonlinear} response to a {\em slowly} varying (compared to
optical period) pump-induced electric field. On the other hand, for the GaAs
type interface, electrons and holes are separated by several atomic layers, so
the corresponding susceptibility $\chi^{(3)}$ is bulk-like. Accordingly, the
EFISHG signal intensities for sample 4 and for GaAs/GaSb heterostructure are
comparable (Fig. 2).  Note that the carriers are more strongly confined at the
"thin" GaSb/InAs interface so we observe a larger EFISHG signal constant
background due to a residual electric field for sample with InSb than with
GaAs type interface.

Finally, let us mention another possible contribution to the EFISHG signal.
The usual description (\ref{pol}) relies on the dipole approximation for
induced electric fields which applies if spatial dependence of ${\cal E}(z,t)$
is smooth. This approximation is standard for Si/SiO interface where the
electric field extends over several atomic layers, and should also be
applicable to semiconductor heterostructures with similar interface size. The
{\em bulk} quadrupole terms, originating from the spatial variation of the SHG
electric field, are smaller than dipole terms and can be
neglected\cite{downer99}. However, in 
the case of GaSb/InAs (with InSb bonds) interface, the electric field strongly
changes on scale of a {\em single} atomic layer, so that {\em induced}
quadrupole terms of the form $\chi^{(Q)} \nabla_z{\cal E}(z,t)[E(\omega)]^2$
should be, in principle, included in the nonlinear polarization (\ref{pol}).
The available experimental data do not allow us to estimate the importance of
such terms.

In summary, we have studied ultrafast dynamics of interfacial electric fields
in GaAs/GaSb and GaAs/GaSb/InAs heterostructures using a pump-probe SHG
technique. We observed a complex evolution of the interfacial fields
originating from the redistribution of carriers between the interfaces. We
also found a strong enhancement of the SHG signal caused by an interband
mixing at the GaSb/InAs interface.  The ability of the EFISHG signal to
monitor spatially separated regions makes pump-probe SHG a unique tool for
studying relaxation and transport phenomena in multilayer semiconductor
structures.

This work was supported by ONR and by the DARPA/SPINS Program.

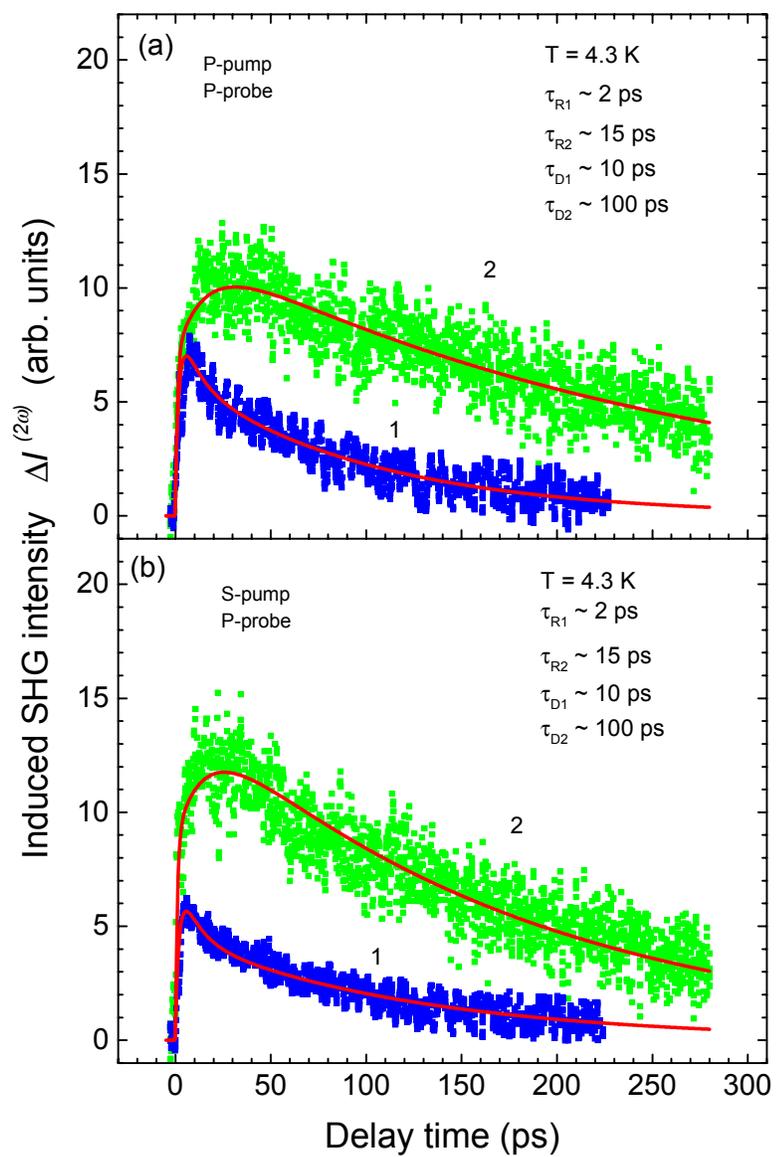
\begin{figure}
\caption{EFISHG signals from GaAs/GaSb sample (blue) and GaAs/GaSb/InAs
sample with InSb type interface (green) measured with (a)
$p$-polarized pump light and (b) $s$-polarized pump light. The fits
with rise/decay exonential function for GaAs/GaSb/InAs (upper curve)
and GaAs/GaSb (lower curve) samples are shown by solid lines.
}
\label{fig:1}
\end{figure}

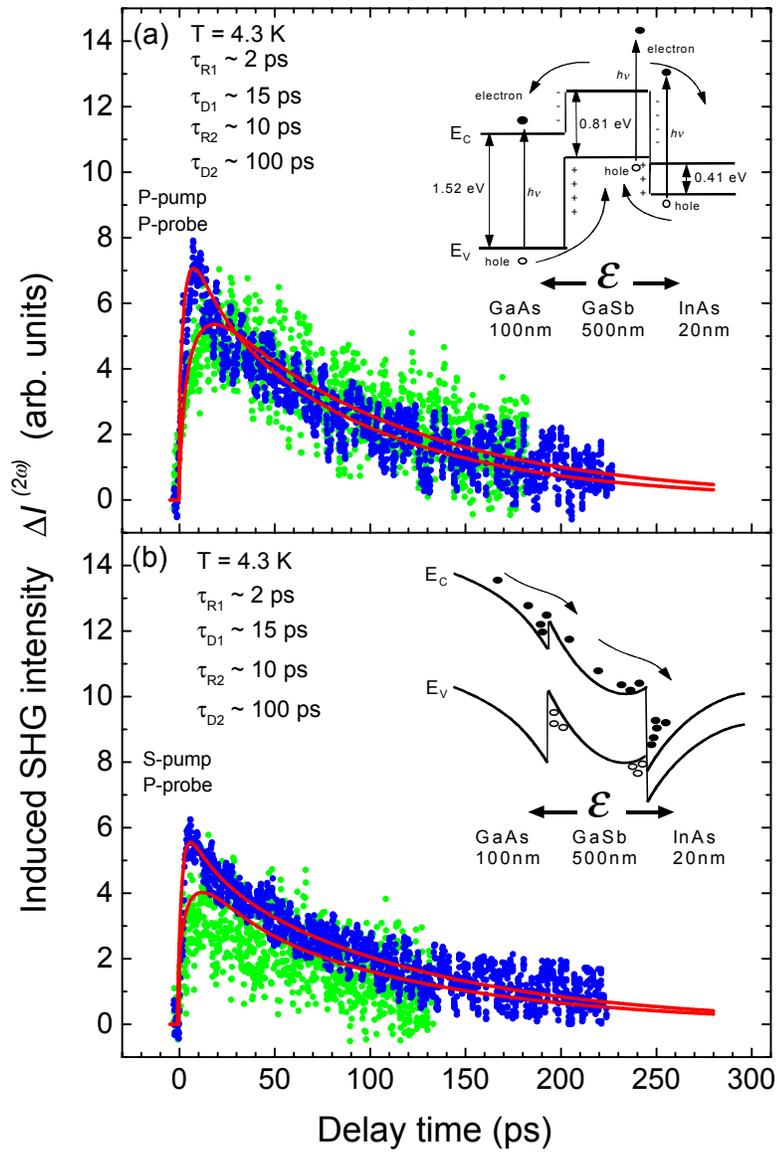
\begin{figure}
\caption{EFISHG signals from GaAs/GaSb sample (blue) and GaAs/GaSb/InAs
sample with GaAs type interface (green) measured with (a)
$p$-polarized pump light and (b) $s$-polarized pump light. The fits
with rise/decay exonential function for GaAs/GaSb/InAs (lower curve)
and GaAs/GaSb (upper curve) samples are shown by solid lines.
The initial band alignment for GaAs/GaSb/InAs 
heterostructure and its realignment due to induced interfacial
electric fields are shown as insets in (a) and (b), respectively.}
\label{fig:2}
\end{figure}


\begin{references}

\bibitem{chemla01}See, e.g., D. S. Chemla and J. Shah,
Nature {\bf 411}, 549 (2001).

\bibitem{perakis00}See, e.g., I. E. Perakis and T. V. Shahbazyan,
Surf. Sci. Reports {\bf 40}, 1 (2000). 

\bibitem{othonos98}See, e.g., A. Othonos,
J. Appl. Phys. {\bf 83}, 1789 (1998).

\bibitem{bolton98} S. Bolton, G. Sucha, D. Chemla, D. L. Sivco, and A. Y. Cho, 
Phys. Rev. B  {\bf 58}, 16326 (1998).

\bibitem{borri01} P. Borri, S. Schneider, W. Langbein, U. Woggon,
A. E. Zhukov, V. M. Ustinov, N. N. Ledentsov, Zh. I. Alferov,
D. Ouyang, and D. Bimberg, 
Appl. Phys. Lett. {\bf 79}, 2633 (2001).

\bibitem{lin87} W. Z. Lin, L. G. Fujimoto, E. P. Ippen, and R. A. Logan, 
Appl. Phys. Lett.  {\bf 50}, 124 (1987).

\bibitem{koshihara97}S. Koshihara, A. Oiwa, M. Hirasawa, S. Katsumoto,
Y. Iye, C. Urano, H. Takagi, and H. Munekata,
Phys. Rev. Lett. {\bf 78}, 4617 (1997).

\bibitem{shen84} Y. R. Shen, {\em The Principles of Nonlinear Optics}
(Wiley, New York, 1984). 


\bibitem{lupke99}See, e.g., Lupke, 
Surf. Sci. Rep. {\bf 35}, 75 (1999), and references therein.

\bibitem{downer99}O. A. Aktisperov, A. A. Fedyanin, A. V. Melnikov,
E. D. Mishina, A. N. Rubtsov, M. H. Anderson, P. T. Wilson, M. ter Beek,
X. F. Hu, J. I. Dadap, and M. C. Downer,
Phys. Rev. B {\bf 60} 8924 (1999).


\bibitem{glinka02} Yu. D. Glinka, W. Wang, S. K. Singh, Z. Marka,
S. N. Rashkeev, Y. Rogachyova, R. Albridge, S. T. Pantelides, N. H. Tolk, 
and G. Lukovsky, 
Phys. Rev. B 65 15 May (2002).

\bibitem{tommasi95} R. Tommasi, P. Langot, and F. Vallee,
Appl. Phys. Lett. {\bf 66}, 1361 (1995). 


\bibitem{qi95} J. Qi, W. Angerer, M. S. Yaganeh, A. G. Yodh, and W. M. Theis, 
Phys. Rev. B {\bf 51}, 13533 (1995).


\bibitem{meyer}I. Vurgaftman, J. R. Meyer, and L. R. Ram-Mohan,
J. Appl. Phys. {\bf 89}, 5815 (2001).

\bibitem{yang97} M. J. Yang, C. H. Yang, B. R. Bennett, and B. V. Shanabrook,
Phys. Rev.  Lett. {\bf 78}, 4613 (1997).

\bibitem{kono97}J. Kono, B. D. McCombe, J.-P. Cheng, I. Lo, W. C. Mitchel, 
and C. E. Stutz,
Phys. Rev. B {\bf 55}, 1617 (1997).

\bibitem{marlow99} T. P. Marlow, L. J. Cooper, D. D. Arnone, N. K. Patel,
D. M. Whittaker, E. H. Linfield, D. A. Ritchie, and M. Pepper, 
Phys. Rev. Lett. {\bf 82}, 2362 (1999).

\bibitem{comanescu02}G. Comanescu, R. J. Wagner, B. D. McCombe,
B. V. Shanabrook, B. R. Bennett, S. K. Singh, J. G. Tischler, and
B. A. Weinstein, 
Physica E (in press) (2002).

\bibitem{bartels97}G. Bartels, G. C. Cho, T. Dekorsy, H. Kurz, A. Stahl, 
and K. K\"{o}hler, 
Phys. Rev. B {\bf 55} 16404 (1997).


%\bibitem{walle89} Van de Walle, Phys. Rev. B 39, 1871 (1989).

%\bibitem{glinka-APL}Yu. D. Glinka, T. V. Shahbazyan, I. E. Perakis,
%N. H. Tolk,X . Liu, Y. Sasaki, and J. K. Furdyna, to be published.
\end{references}
\end{document}